	\documentclass[showpacs,preprintnumbers,amsmath,amssymb,prl, superscriptaddress, nofootinbib, aps,prl,superscriptaddress]{revtex4-1}
	
	\usepackage{lineno}
	\usepackage{graphicx}
	\usepackage{dcolumn}
	\usepackage{bm}
	\usepackage{amsmath}
	\usepackage{amssymb} 
	\usepackage{relsize}
	
	\usepackage{array}
	\usepackage[caption=false]{subfig}
	\usepackage{hyperref}

\begin{document}
	
	\title{Constraints on ultra-high-energy cosmic ray sources from a search for neutrinos above 10 PeV with IceCube}
	\date{\today}
	
	\affiliation{III. Physikalisches Institut, RWTH Aachen University, D-52056 Aachen, Germany}
	\affiliation{Department of Physics, University of Adelaide, Adelaide, 5005, Australia}
	\affiliation{Dept.~of Physics and Astronomy, University of Alaska Anchorage, 3211 Providence Dr., Anchorage, AK 99508, USA}
	\affiliation{CTSPS, Clark-Atlanta University, Atlanta, GA 30314, USA}
	\affiliation{School of Physics and Center for Relativistic Astrophysics, Georgia Institute of Technology, Atlanta, GA 30332, USA}
	\affiliation{Dept.~of Physics, Southern University, Baton Rouge, LA 70813, USA}
	\affiliation{Dept.~of Physics, University of California, Berkeley, CA 94720, USA}
	\affiliation{Lawrence Berkeley National Laboratory, Berkeley, CA 94720, USA}
	\affiliation{Institut f\"ur Physik, Humboldt-Universit\"at zu Berlin, D-12489 Berlin, Germany}
	\affiliation{Fakult\"at f\"ur Physik \& Astronomie, Ruhr-Universit\"at Bochum, D-44780 Bochum, Germany}
	\affiliation{Physikalisches Institut, Universit\"at Bonn, Nussallee 12, D-53115 Bonn, Germany}
	\affiliation{Universit\'e Libre de Bruxelles, Science Faculty CP230, B-1050 Brussels, Belgium}
	\affiliation{Vrije Universiteit Brussel, Dienst ELEM, B-1050 Brussels, Belgium}
	\affiliation{Dept.~of Physics, Massachusetts Institute of Technology, Cambridge, MA 02139, USA}
	\affiliation{Dept.~of Physics and Institute for Global Prominent Research, Chiba University, Chiba 263-8522, Japan}
	\affiliation{Dept.~of Physics and Astronomy, University of Canterbury, Private Bag 4800, Christchurch, New Zealand}
	\affiliation{Dept.~of Physics, University of Maryland, College Park, MD 20742, USA}
	\affiliation{Dept.~of Physics and Center for Cosmology and Astro-Particle Physics, Ohio State University, Columbus, OH 43210, USA}
	\affiliation{Dept.~of Astronomy, Ohio State University, Columbus, OH 43210, USA}
	\affiliation{Niels Bohr Institute, University of Copenhagen, DK-2100 Copenhagen, Denmark}
	\affiliation{Dept.~of Physics, TU Dortmund University, D-44221 Dortmund, Germany}
	\affiliation{Dept.~of Physics and Astronomy, Michigan State University, East Lansing, MI 48824, USA}
	\affiliation{Dept.~of Physics, University of Alberta, Edmonton, Alberta, Canada T6G 2E1}
	\affiliation{Erlangen Centre for Astroparticle Physics, Friedrich-Alexander-Universit\"at Erlangen-N\"urnberg, D-91058 Erlangen, Germany}
	\affiliation{D\'epartement de physique nucl\'eaire et corpusculaire, Universit\'e de Gen\`eve, CH-1211 Gen\`eve, Switzerland}
	\affiliation{Dept.~of Physics and Astronomy, University of Gent, B-9000 Gent, Belgium}
	\affiliation{Dept.~of Physics and Astronomy, University of California, Irvine, CA 92697, USA}
	\affiliation{Dept.~of Physics and Astronomy, University of Kansas, Lawrence, KS 66045, USA}
	\affiliation{Dept.~of Astronomy, University of Wisconsin, Madison, WI 53706, USA}
	\affiliation{Dept.~of Physics and Wisconsin IceCube Particle Astrophysics Center, University of Wisconsin, Madison, WI 53706, USA}
	\affiliation{Institute of Physics, University of Mainz, Staudinger Weg 7, D-55099 Mainz, Germany}
	\affiliation{Department of Physics, Marquette University, Milwaukee, WI, 53201, USA}
	\affiliation{Universit\'e de Mons, 7000 Mons, Belgium}
	\affiliation{National Research Nuclear University MEPhI (Moscow Engineering Physics Institute), Moscow, Russia}
	\affiliation{Physik-department, Technische Universit\"at M\"unchen, D-85748 Garching, Germany}
	\affiliation{Institut f\"ur Kernphysik, Westf\"alische Wilhelms-Universit\"at M\"unster, D-48149 M\"unster, Germany}
	\affiliation{Bartol Research Institute and Dept.~of Physics and Astronomy, University of Delaware, Newark, DE 19716, USA}
	\affiliation{Dept.~of Physics, Yale University, New Haven, CT 06520, USA}
	\affiliation{Dept.~of Physics, University of Oxford, 1 Keble Road, Oxford OX1 3NP, UK}
	\affiliation{Dept.~of Physics, Drexel University, 3141 Chestnut Street, Philadelphia, PA 19104, USA}
	\affiliation{Physics Department, South Dakota School of Mines and Technology, Rapid City, SD 57701, USA}
	\affiliation{Dept.~of Physics, University of Wisconsin, River Falls, WI 54022, USA}
	\affiliation{Oskar Klein Centre and Dept.~of Physics, Stockholm University, SE-10691 Stockholm, Sweden}
	\affiliation{Dept.~of Physics and Astronomy, Stony Brook University, Stony Brook, NY 11794-3800, USA}
	\affiliation{Dept.~of Physics, Sungkyunkwan University, Suwon 440-746, Korea}
	\affiliation{Dept.~of Physics, University of Toronto, Toronto, Ontario, Canada, M5S 1A7}
	\affiliation{Dept.~of Physics and Astronomy, University of Alabama, Tuscaloosa, AL 35487, USA}
	\affiliation{Dept.~of Astronomy and Astrophysics, Pennsylvania State University, University Park, PA 16802, USA}
	\affiliation{Dept.~of Physics, Pennsylvania State University, University Park, PA 16802, USA}
	\affiliation{Dept.~of Physics and Astronomy, University of Rochester, Rochester, NY 14627, USA}
	\affiliation{Dept.~of Physics and Astronomy, Uppsala University, Box 516, S-75120 Uppsala, Sweden}
	\affiliation{Dept.~of Physics, University of Wuppertal, D-42119 Wuppertal, Germany}
	\affiliation{DESY, D-15735 Zeuthen, Germany}
	
	\author{M.~G.~Aartsen}
	\affiliation{Department of Physics, University of Adelaide, Adelaide, 5005, Australia}
	\author{K.~Abraham}
	\affiliation{Physik-department, Technische Universit\"at M\"unchen, D-85748 Garching, Germany}
	\author{M.~Ackermann}
	\affiliation{DESY, D-15735 Zeuthen, Germany}
	\author{J.~Adams}
	\affiliation{Dept.~of Physics and Astronomy, University of Canterbury, Private Bag 4800, Christchurch, New Zealand}
	\author{J.~A.~Aguilar}
	\affiliation{Universit\'e Libre de Bruxelles, Science Faculty CP230, B-1050 Brussels, Belgium}
	\author{M.~Ahlers}
	\affiliation{Dept.~of Physics and Wisconsin IceCube Particle Astrophysics Center, University of Wisconsin, Madison, WI 53706, USA}
	\author{M.~Ahrens}
	\affiliation{Oskar Klein Centre and Dept.~of Physics, Stockholm University, SE-10691 Stockholm, Sweden}
	\author{D.~Altmann}
	\affiliation{Erlangen Centre for Astroparticle Physics, Friedrich-Alexander-Universit\"at Erlangen-N\"urnberg, D-91058 Erlangen, Germany}
	\author{K.~Andeen}
	\affiliation{Department of Physics, Marquette University, Milwaukee, WI, 53201, USA}
	\author{T.~Anderson}
	\affiliation{Dept.~of Physics, Pennsylvania State University, University Park, PA 16802, USA}
% ...
\collaboration{IceCube Collaboration}
\noaffiliation

	\begin{abstract}
		We report constraints on the sources of ultra-high-energy cosmic rays (UHECRs) above $10^{9}$\;GeV, based on an analysis of seven years of 
		IceCube data. 
		This analysis efficiently selects very high energy neutrino-induced events which have deposited energies from $5 \times 10^5$ GeV to above $10^{11}$\;GeV.  
		Two neutrino-induced events with an estimated deposited energy of $(2.6 \pm 0.3) \times 10^6$\;GeV, the highest neutrino energy observed so far, and $(7.7 \pm 2.0) \times 10^5$\;GeV were detected. 
		The atmospheric background-only hypothesis of detecting these events is rejected at 3.6$\sigma$.
		The hypothesis that the observed events are of cosmogenic origin is also rejected at $>$99\% CL because of the limited deposited energy and the non-observation of events at higher energy, while their observation is consistent with an astrophysical origin.
		Our limits on cosmogenic neutrino fluxes disfavor the UHECR sources having cosmological evolution stronger than the star formation rate, e.g., active galactic nuclei and $\gamma$-ray bursts, assuming proton-dominated UHECRs. 
		Constraints on UHECR sources including mixed and heavy UHECR compositions are obtained for models of neutrino production within UHECR sources.
		Our limit disfavors a significant part of parameter space for active galactic nuclei and new-born pulsar models.
		These limits on the ultra-high-energy neutrino flux models are the most stringent to date.
	\end{abstract}
	
	\maketitle
	\paragraph{Introduction ---}
	The sources of ultra-high-energy cosmic rays (UHECRs; cosmic-ray energy $E_{CR}$ $\gtrsim$ $10^{18}$\;eV) 
	remain unidentified \cite{pdg_cr}.
	The majority of the candidate objects are extra-Galactic, such as Active 
	Galactic Nuclei (AGN) \cite{AGN_Rachen1993,AGN_Mannheim1995,AGN_Atoyan2001,AGN_Atoyan2003,AGN_AlvarezMuniz2004} , $\gamma$-ray bursts \cite{GRB_Vietri1995,GRB_Waxman1995,GRB_Waxman1997,GRB_Waxman2000,GRB_Dermer2002,GRB_Globus2015,GRB_Bustamante2015}, and starburst galaxies \cite{Starburst_Loeb2006,Starburst_Thompson2006,Starburst_Murase2013,Starburst_Tamborra2014,Starburst_Wang2014,Starburst_Emig2015}. 
	UHECR interactions with ambient photons and matter at sources generate astrophysical neutrinos with 5\% of 
	the parent UHECR energy, on average \cite{Sophia_Mucke2000,Flavor_Lipari2007,Photomeson_Yoshida2014}. 
	Thus, a substantial fraction of these Extremely High-Energy (EHE) astrophysical neutrinos is expected to have an energy above $10^7$ GeV. 
	Moreover, neutrinos with energies above $\sim$$10^{7}$~GeV are expected to be produced in the interactions between the highest-energy 
	cosmic rays and background photons in the universe~\cite{gzk}. In the following we refer to the neutrinos produced in these interactions as cosmogenic neutrinos~\cite{berezinsky69}.
	These astrophysical and cosmogenic EHE neutrinos can constitute key messengers identifying currently unknown 
	cosmic accelerators, possibly in the distant universe, because their propagation is not influenced by background photon or magnetic fields.
	
	In this Letter, we report results of an analysis of seven years of IceCube data obtained in the search for  
	diffuse neutrinos with energies larger than $5 \times 10^5$ GeV.
	The current analysis is optimized in particular for the neutrinos with energies above $10^7$ GeV, which is higher in energy than the other IceCube analyses \cite{upmu, HESE3}. 
	The analysis described here is based on data taken between April 2008 and May 2015, corresponding to 2426\;days of effective 
	livetime. This is approximately three times more data than the previous IceCube 
	EHE neutrino search based on two years of data~\cite{EHE2012}. No cosmogenic neutrino candidate was observed 
	in that study, but two PeV events were detected~\cite{EHEPRL}. Stringent limits were placed on 
	cosmogenic neutrino fluxes, and it was shown that astrophysical objects with emission rates per comoving 
	space density as a function of redshift $\varPsi_s(z)$ following a strong cosmological evolution, such as 
	Fanaroff-Riley type-II (FRII) radio galaxies, are disfavored as highest-energy cosmic-ray sources.

	\paragraph{Data selection and analysis ---}
	IceCube is a cubic-kilometer deep-underground Cherenkov neutrino detector located at the South Pole~\cite{icecube}, 
	which is designed to measure neutrinos with energies above $10^2$ GeV. The construction of the IceCube detector was completed in December 2010. 
	The array comprises 5160 optical sensors~\cite{DOM, pmt}  on 86 vertical strings distributed over a 1-km$^3$ 
	instrumented ice volume at 1450--2450 m depth. Additional particle shower sensors at the surface constitute the IceTop air 
	shower array \cite{icetop}. In 2008--2009, 2009--2010, and 2010--2011, 40, 59, and 79 strings out of 86 were operational.
	Since 2011, IceCube has been recording data using the completed array.
	
	The primary background in this analysis consists of downward-going muon bundles composed of a 
	large number of muons produced in cosmic-ray interactions in the atmosphere. This background was 
	simulated using the {\sc corsika} package~\cite{corsika} with the {\sc sibyll}~\cite{sibyll} hadronic 
	interaction model. The secondary background is atmospheric neutrinos produced by the decay of charged mesons from 
	cosmic-ray interactions in the atmosphere.
	The atmospheric neutrino simulation was generated by the IceCube {neutrino-generator} program based on 
	the {\sc anis} code~\cite{anis}.
	At energies above $\sim$$10^{6}$~GeV, prompt atmospheric neutrinos from short-lived heavy meson decays are expected to dominate over 
	conventional atmospheric neutrinos from pion and kaon decays.
	While a flux of prompt atmospheric neutrinos must exist, it has not been experimentally observed. 
	The conventional atmospheric neutrino model from \cite{Honda2007}, and the prompt model 
	presented in \cite{prompt} both incorporating the cosmic-ray knee model given in \cite{TG} are included in the 
	background estimation. An updated calculation \cite{BERSS} of the prompt flux \cite{prompt} predicts a reduced 
	prompt flux by a factor of $\sim$2. 
	The experimental data agree well with lower energy background predictions \cite{EHE2012}. 
	Cosmogenic and astrophysical neutrinos are simulated using the {\sc juli}e{\sc t} 
	package~\cite{juliet} as in our earlier work~\cite{EHE2012}.
	
\begin{figure}[t]
			\begin{center}
			\includegraphics[width=3.2in]{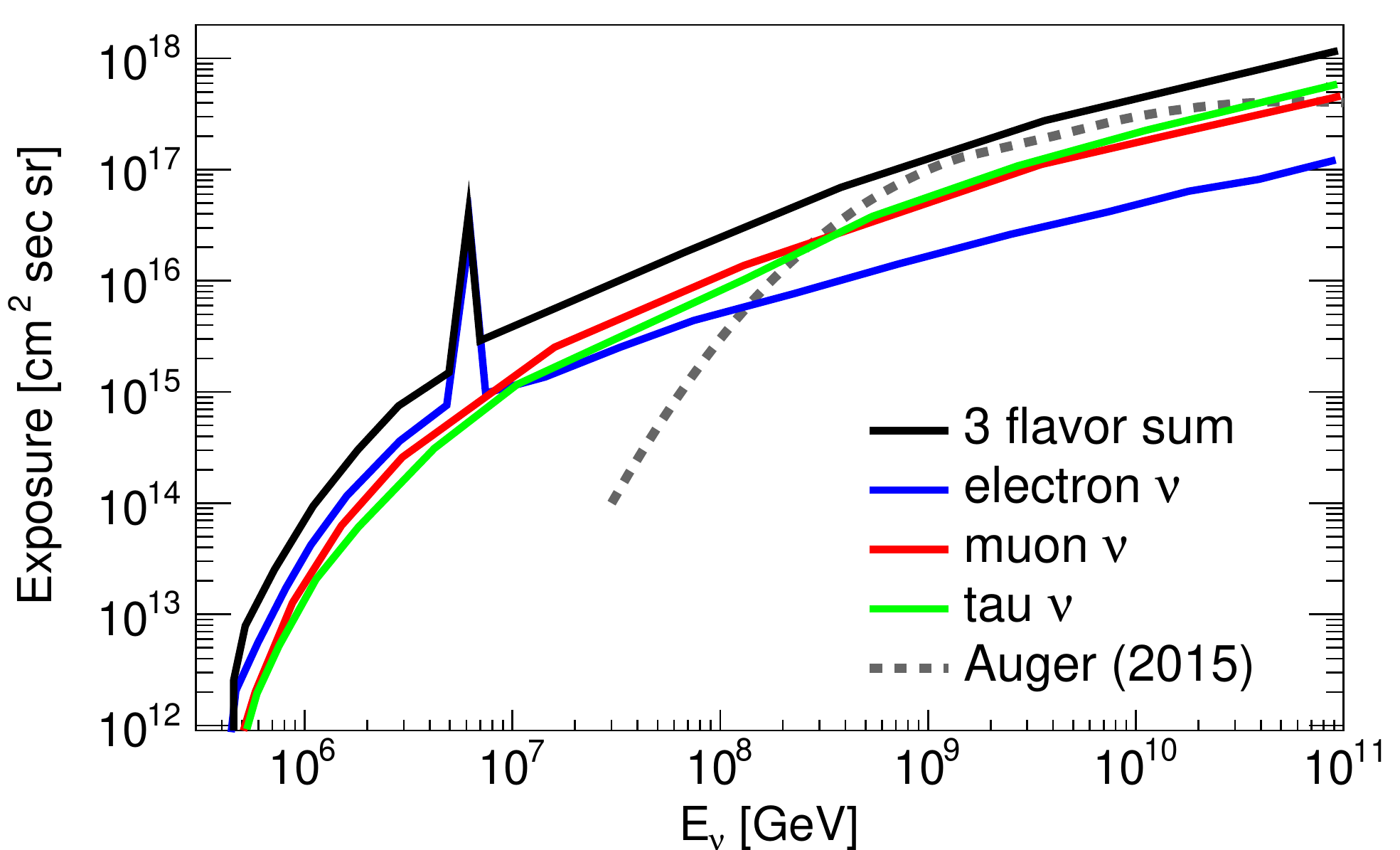}
			\end{center}
			\caption{(Color online)Solid angle and time integrated exposure from April 2008 through May 2015 as a function of neutrino 
				energy by flavor and flavor sum, including neutrino absorption effects in the Earth. 
				The sharp peaked structure at $6.3\times10^6$\;GeV for electron neutrinos is due to the Glashow resonance \cite{glashow}.
				The Auger exposure\;\cite{auger15} is included for reference. 
			}
			\label{fig:effarea}
		\end{figure}

	The majority of atmospheric backgrounds deposit less energy in IceCube than 
	the EHE neutrino signal.  We reject most of the background by cutting 
	events with low energy deposition.
	The number of observed Cherenkov photons is used as a proxy for the deposited energy. 
	The majority of the background is removed by requiring that the measured Number of PhotoElectrons (NPE) 
	is larger than a zenith angle-dependent threshold. 
	The reconstructed zenith angle is obtained using a $\chi^2$ fit to a simple track hypothesis~\cite{iLF}.
	The quality of the reconstruction is evaluated via the $\chi^2_{\rm track}/ndf$ where $ndf$ is the number of degrees of freedom.
	The selection threshold is optimized for the cosmological neutrino model\;\cite{ahlers2010} and kept constant for 
	each detector configuration. The criteria are qualitatively equivalent to those used in~\cite{EHE2012}, with 
	details given in the Supplemental Material \cite{supple}.
	Events with more than a single IceTop hit in the time interval of $-1 \mu$s $\leq$ $t_{ca}$ $\leq$ $1.5 \mu$s are rejected. Here, $t_{ca}$ is the time when the reconstructed downward-going track is at  
	the closest approach to the IceTop optical sensors. 
	The exposure of this analysis for each neutrino flavor is shown in 
	Fig.\;\ref{fig:effarea} along with the summed exposure.	
		
	The background event rate induced by the atmospheric muons and 
	neutrinos is reduced from $\sim$2.8 kHz trigger-level rate to $0.064^{+0.023}_{-0.039}$ events per 2426 days of livetime.
	The expected event rates for cosmogenic and astrophysical models are shown in Tables\;\ref{table:CL} and \ref{table:CL2}, respectively.
	Only electron and muon neutrinos are produced when UHECRs interact with photons or matter.
	As a result of flavor oscillation, $\nu_e:\nu_\mu:\nu_\tau$ = 1:1:1 on Earth, assuming the standard full pion decay chain of neutrino production \cite{Flavor_Barenboim2003,Flavor_Lipari2007,Flavor_Bustamante2015}. This is compatible with TeV-PeV flavor ratio measurements \cite{IceCube_FlavorPRL,IceCube_Globalfit,Flavor_Palladino2015,HESE4Fit_Vincent}. The neutrino distributions are summed 
	over the three flavors. Equal neutrino and anti-neutrino fluxes, indistinguishable in IceCube, are assumed.
		%
	
	Two events were observed in the present 2426-day IceCube sample.
	The best estimates of the deposited energy are (7.7 $\pm$ 2.0) $\times$ $10^5$\;GeV and (2.6 $\pm$ 0.3) $\times$ $10^6$\;GeV, 
	in the form of a spherical particle shower and an upward-moving track at a zenith angle of $101^\circ$ \cite{atel}, 
	respectively. 
	Three previously observed PeV events~\cite{EHEPRL, HESEPRL} do not pass the current event 
	selection, due to the increased NPE threshold for events with 
	$\chi^2_{\rm track}/ndf \geq 80$.
	
	The sample satisfying the selection criteria is analyzed using the binned Poisson Log-likelihood ratio (LLR) method \cite{LLHtest}. 
	The events are binned in both the reconstructed zenith angle and the energy proxy, with only the energy proxy information 
	being used for events with large $\chi^2_{\rm track}/ndf$.
	The zenith angle and energy proxy used in the LLR test are the results of refined reconstruction using a maximum 
	likelihood method\;\cite{LLH, Energy}.  
	The hypothesis that the two observed events are of atmospheric origin is tested using an ensemble of pseudo-experiment 
	trials to derive the LLR test statistic distribution. The test rejects the atmospheric background-only 
	hypothesis with a p-value of $0.014$\% (3.6$\sigma$). Furthermore, the hypothesis that the two events are of cosmogenic 
	origin is rejected with a p-value of 0.3\%, because of the low observed deposited energy and the absence of detected events at higher 
	energy. However, the observations are compatible with a generic astrophysical $E^{-2}$ power-law flux with a p-value of 92.3\%.
	The energy deposited and the zenith angles of the two observed events are better described by a neutrino spectrum
	softer than the spectrum of $\geq 10^8$\;GeV neutrinos, which experience strong absorption effects during their propagation through the Earth.
	This observation allows us to set an upper limit on a neutrino flux extending above $10^7$\;GeV. The limits also are derived using the LLR method. 
	Cosmogenic neutrino models are tested by adding an unbroken $E^{-2}$ flux without cutoff as a nuisance parameter to explain the observed two events.
	
	The systematic uncertainties are estimated similarly to the previous publication\;\cite{EHE2012}. The primary sources of 
	uncertainty are simulations of the detector responses and optical properties of the ice.
	These uncertainties are evaluated with {\it in-situ} 
	calibration system using a light source and optical sensor sensitivity studies in the laboratory.
	Uncertainties of $^{+13\%}_{-42\%}$ and $^{+2\%}_{-7\%}$ are estimated for the number of background and signal events, 
	respectively. 
	In addition, uncertainties of $-11\%$ are introduced to the
	neutrino-interaction cross-section based on {\sc cteq5} \cite{cteq5} calculated as \cite{gqrs} and $+10\%$ by the
	photonuclear energy losses \cite{photonuclear}. The uncertainty on the neutrino-interaction cross section is
	from \cite{csms_11}. The uncertainty associated with photonuclear cross section is estimated by comparing the current calculation with the soft-component-only model.
	Uncertainty of $^{+34\%}_{-44\%}$ associated with the 
	atmospheric background is also included. The error is dominated by the experimental uncertainty of CR spectrum measurements ($\pm$30\%) \cite{nagano, pdg_cr},
	theoretical uncertainty on the prompt flux calculation \cite{prompt}, and the primary CR composition.
	All the resultant limits presented in this Letter include systematic uncertainties. 
	Taking the maximally and minimally estimated background and signal distributions in a 1$\sigma$ error range by adding 
	systematic uncertainties in quadrature, each signal and background combination results in an upper limit. The weakest 
	limit is taken as a conservative upper limit including systematic uncertainties.
	The uncertainty is energy-dependent and, thus, it is model-spectrum-shape dependent.
	Model-dependent limits are generally weakened by $\sim$20\% and $\sim$30\% for 
	cosmogenic and astrophysical-neutrino models, respectively. 	
	
	\begin{figure}[t]
		\begin{center}
			\includegraphics[width=3.6in]{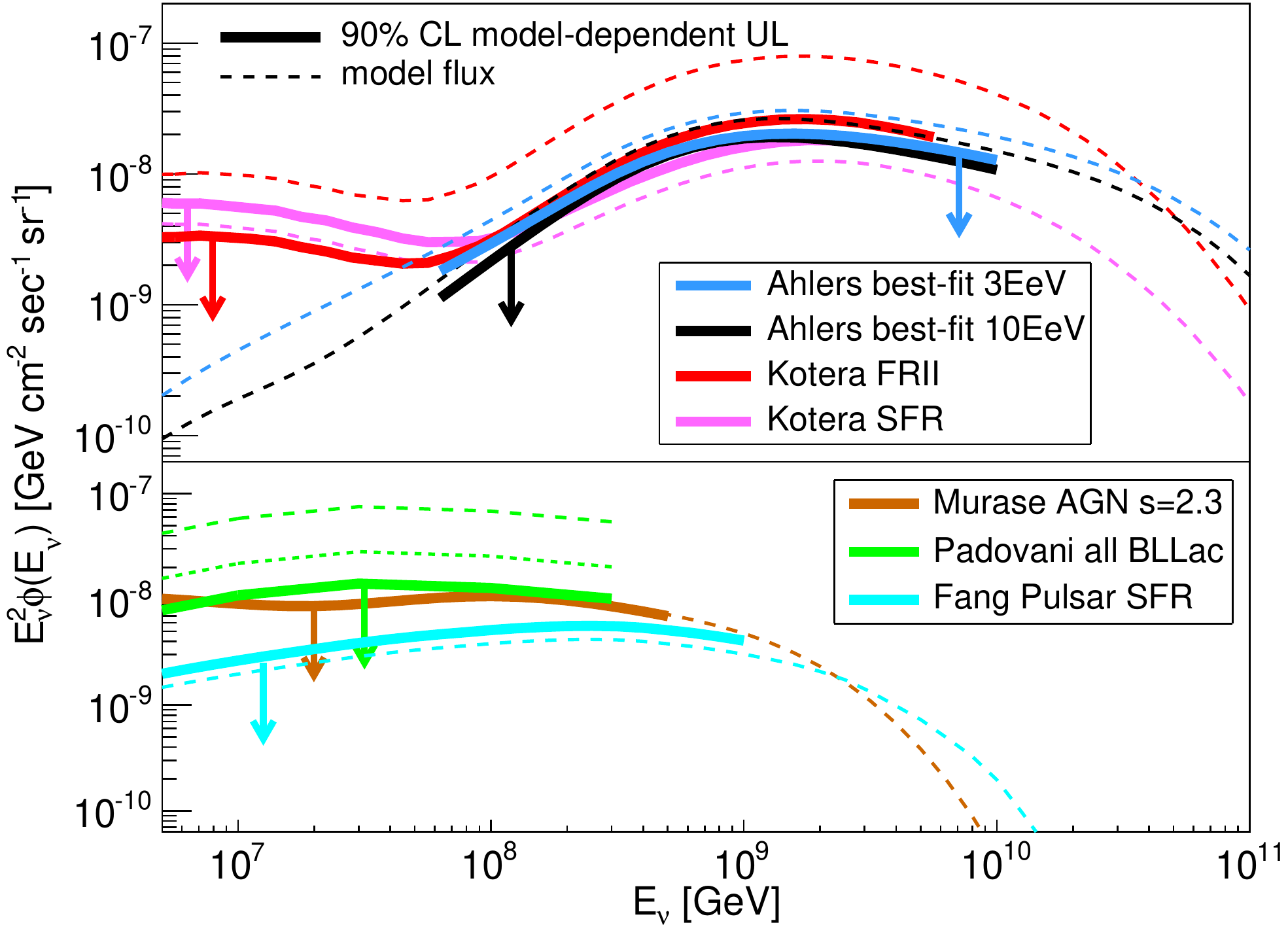}
		\end{center}
		\caption{
			(Color online) Model-dependent 90\% confidence-level limits (solid lines) for (upper panel) proton cosmogenic-neutrino predictions (dashed lines) 
			from Ahlers\;\cite{ahlers2010} and Kotera\;\cite{kotera2010} and (lower panel) astrophysical neutrino fluxes from AGN (BLR) models of Murase\;\cite{murase14} 
			and Padovani (long dashes: $Y_{\nu\gamma}$ = 0.8, short dashes: $Y_{\nu\gamma}$ = 0.3)\;\cite{padovani15}, and Fang pulsar model \;\cite{ke14}. 
			The range of limits indicates the central 90\% energy region.
			Two lines of the Ahlers model represent different threshold energy of the extragalactic UHECR component.
			The deviation of the Kotera and Ahlers models below $10^8$ GeV is due to different models of the extagalactic background light assumed for the calculation. 
			The wide energy coverage of the current analysis (Fig.\;\ref{fig:effarea}) allows a stringent model-dependent limit to be placed for both cosmogenic 
			and astrophysical models.
		}
		\label{fig:MD}
	\end{figure}
		\begin{figure}[hbt]
			\begin{center}
				\includegraphics[height=1.55in]{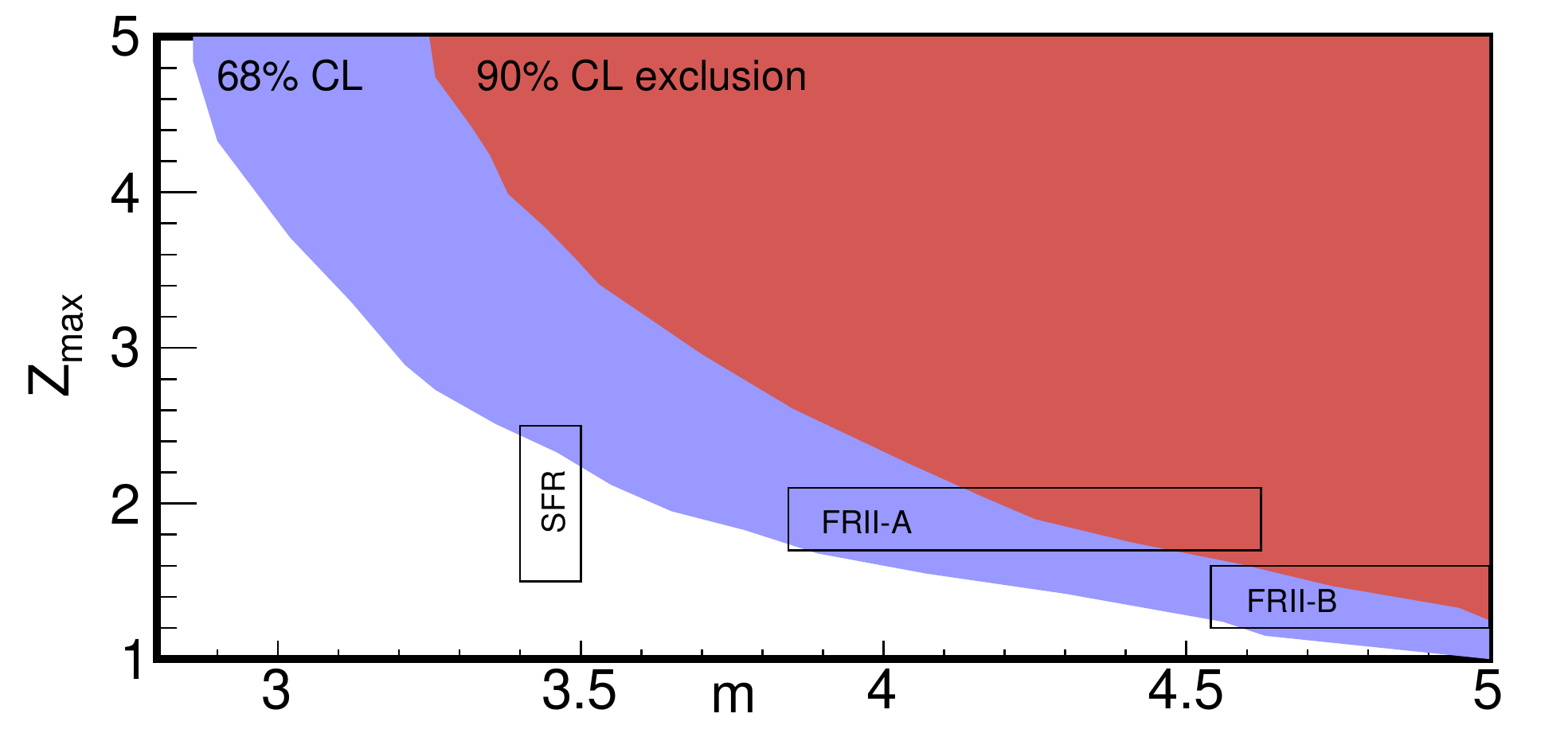}\\
				\includegraphics[height=1.55in]{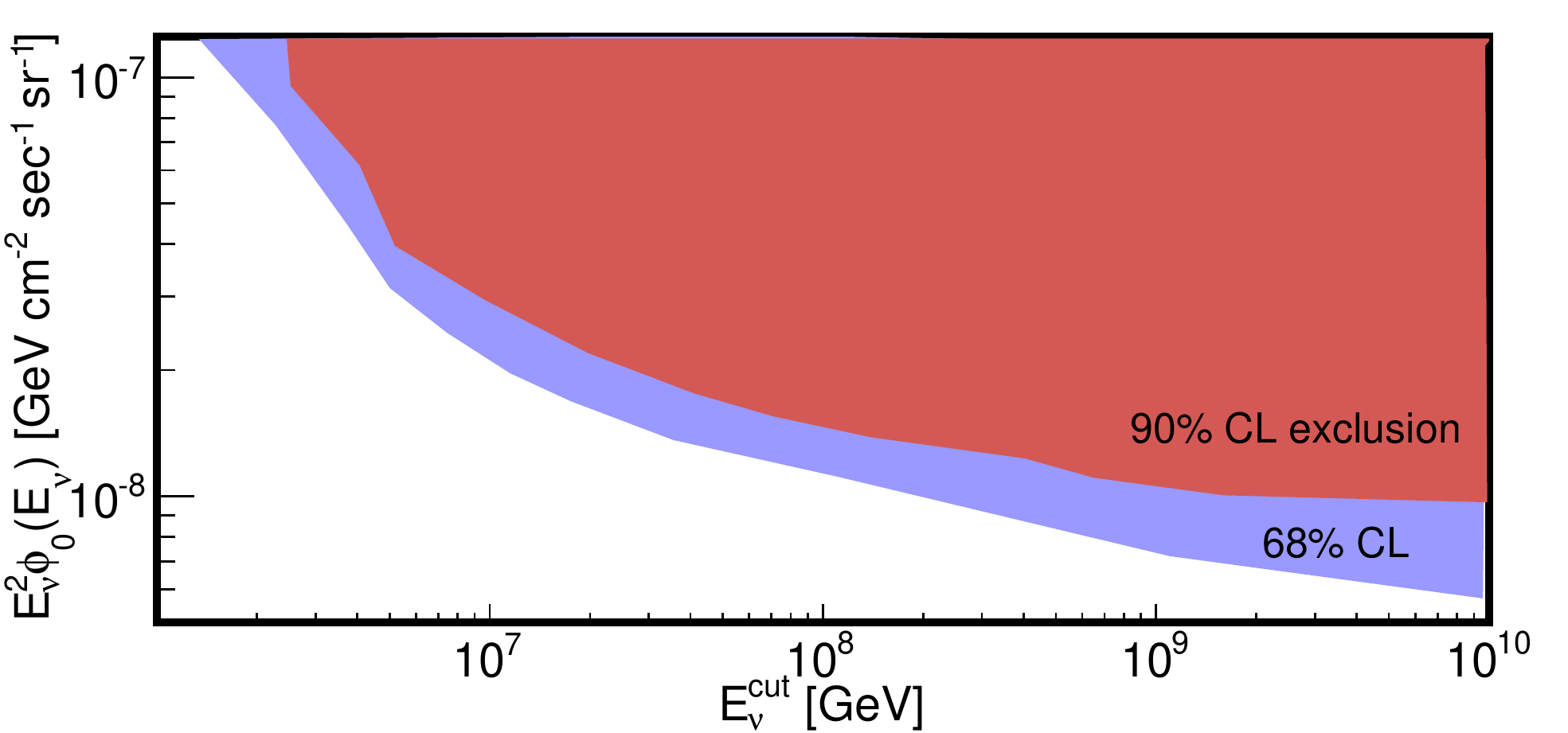}
			\end{center}	
			\caption{(Color online) Constraints on UHECR source evolution model and all flavor $E^{-2}$ power-law 
				flux model parameters. The colored areas represent parameter space excluded by the current analysis.
				(Top) Cosmogenic flux parameters $m$ and $z_{max}$ of UHECR-source cosmological evolution 
				function of the form $\psi_s(z) \propto (1+z)^m$, assuming proton-dominant UHECR primaries with only 
				the CMB as the background photon field.
				A semi-analytic formulation\;\cite{gzkanalytic}, with the injected proton spectrum of $E^{-2.5}$ up to $6 \times 10^{11}$ GeV, is used to estimate the neutrino flux. The boxes indicate 
				approximate parameter regions for SFR\;\cite{SFR} and FRII\;(-A \cite{inoue09} and -B \cite{Ajello12}) neglecting the minor far-redshift contributions.
				(Bottom) 	
				Upper limits on $E^{-2}$ power-law neutrino flux normalization $\phi_0$ and spectral cutoff energy $E^{cut}_{\nu}$. 
			}
			\label{fig:contours}
		\end{figure}
		\begin{figure}[hbt]
			\begin{center} 				
				\includegraphics[width=3.4in]{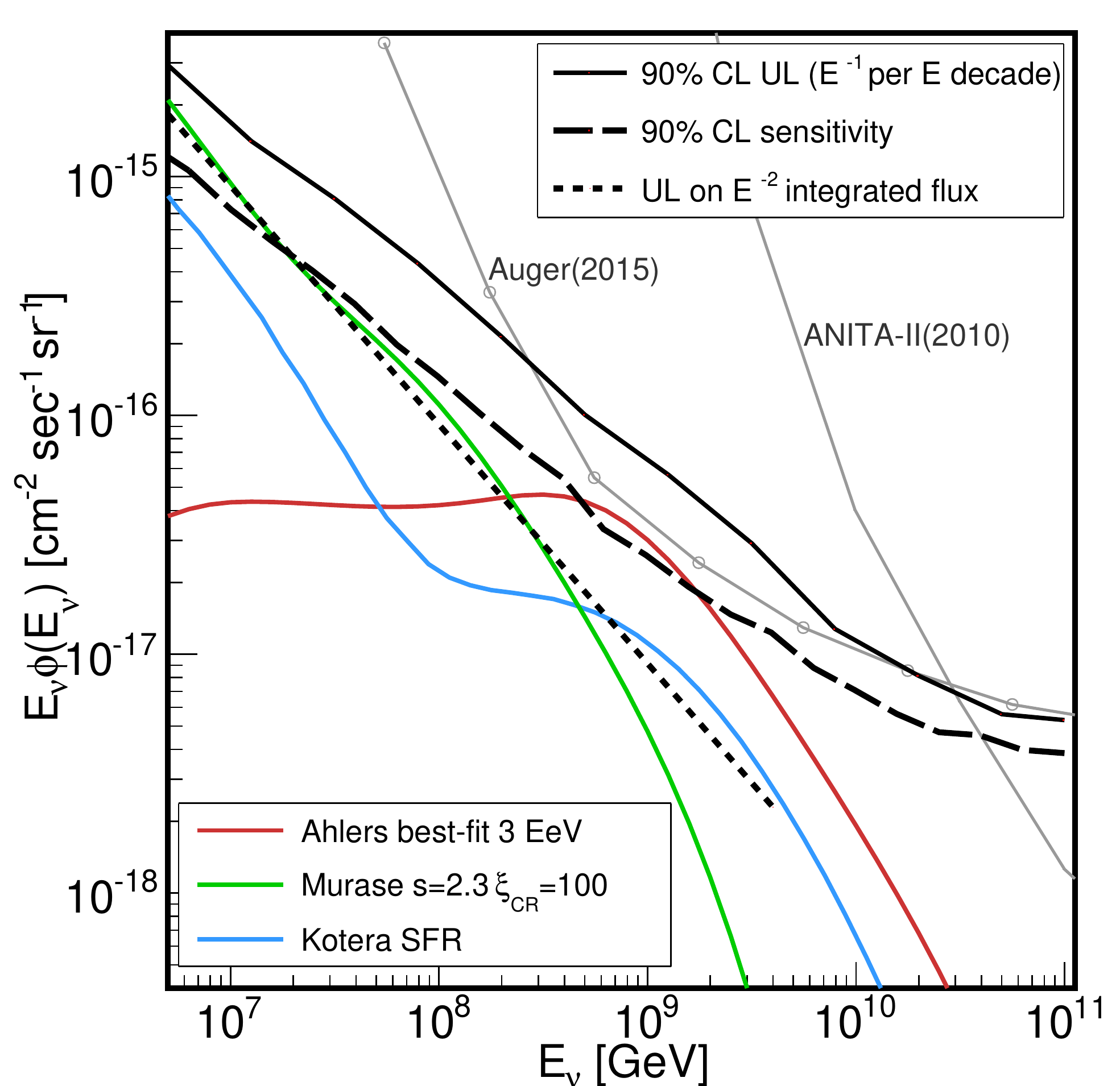}
				\end{center}
				\caption{(Color online) All-flavor-sum neutrino flux quasi-differential 90\%-CL upper limit on one energy decade $E^{-1}$ flux windows (solid line). 
						The limits are derived using a log-likelihood ratio method. The median null observation limit (sensitivity) is also shown (dashed line).
						Cosmogenic-neutrino model predictions (assuming primary protons) are 
						shown for comparison:
						Kotera {\it et al.}~\cite{kotera2010}, 
						Ahlers {\it et al.}~\cite{ahlers2010}, and an astrophysical neutrino model from Murase {\it et al.}~\cite{murase14}.
						Model-independent differential limits on one energy decade $E^{-1}$ flux from Auger~\cite{auger15} and ANITA-II~\cite{anita2} 
						with appropriate 
						normalization are also shown. A model-dependent upper limit on an unbroken $E^{-2}$ power-law flux from the current analysis 
						($E_{\nu}^2\phi < 9.2\times10^{-9}$ GeV/cm$^2$\;s\;sr) is shown for reference (dotted line).
				}
				\label{fig:differential}
		\end{figure}
	\paragraph{Cosmogenic neutrinos ---} 
	We tested cosmogenic neutrino models. Aside from the primary composition dependence, the cosmogenic neutrino 
	rates in the current analysis depend significantly on the UHECR source evolution function that characterize the source 
	classes. Table\;\ref{table:CL} represents the p-values and associated 90\%-CL\ limits for cosmogenic models.
	The models from \cite{ahlers2010} are constructed in such a manner that the cosmogenic $\gamma$-ray emission from the decays of $\pi^0$ produced by the interactions of UHECRs with the cosmic microwave background (CMB) is consistent with the Fermi-LAT measurements of the diffuse extragalactic $\gamma$-ray background\;\cite{fermi2010, fermi2015}.
	Our constraints on these models imply that the majority of the observed $\gamma$-ray background is unlikely to be of cosmogenic origin.
		
	Limits on cosmogenic neutrino models\;\cite{kotera2010, SimProp15} using two classes of source-evolution functions are presented in Table\;\ref{table:CL}. 
	One evolution function is the star formation rate (SFR) \cite{SFR}, which is a generic measure of structure formation history in 
	the universe, and the other is that of FRII radio-loud AGN \cite{inoue09,Ajello12}. 
	The cosmogenic models assuming FRII-type evolution have already been constrained by the previous study\;\cite{EHE2012}. In addition, these strong evolution models may conflict with the observed $\gamma$-ray background \cite{ahlers2010,Berezinsky2011,Berezinsky2016}. The current analysis not only strongly constrains the FRII-type but also is beginning to constrain the parameter space where SFR drives UHECR source evolution.  
	The predicted neutrino spectra and the corresponding model-dependent limits are presented in Fig.\;\ref{fig:MD}. 
	When the primaries are heavy nuclei, photodisintegration is more likely than pion production, hence the flux of cosmogenic muon neutrinos is suppressed \cite{kotera2010,Hooper2005,Roulet2013,Ahlers2012,Ave2015}.
	
	Thus the limit on the proton composition cosmogenic models could also be considered as the limit on the proton fraction of a mixed-composition UHECR model for the 
	given evolution model.   
	
	A more generic scanning of parameter space for the source evolution function, $\varPsi_s(z)\propto (1+z)^m$, up to the maximum source extension 
	in redshift $z\leq z_{\rm max}$, was also performed using an analytical parameterization\;\cite{gzkanalytic}. Because only the CMB is assumed as the 
	target photon field in the parameterization, the limits are systematically weaker than that on the models that include 
	extragalactic background light, such as infrared and optical photons, with the given evolution parameters.
	The resultant exclusion contour is shown in the upper panel of Fig.\;\ref{fig:contours}. Each point represents a given cosmogenic-neutrino 
	model --- normalized by fitting the UHECR spectrum to data \cite{gzkanalytic} --- and the contour represents the exclusion confidence limit calculated using the LLR method. 
	The UHECR spectrum dependence of cosmogenic neutrino model is also studied in \cite{Baerwald}.
	Our results disfavor a large portion of the 
	parameter space where $m$ $\geq3.5$ for sources distributed up to $z_{\rm max}=2$. 
	These constraints imply that the sources of UHECRs seem to evolve more slowly than the SFR. Otherwise, a proton-dominant composition at the highest energies, in particular the dip model\;\cite{dip}, is excluded \cite{Ahlers2009}, as studied also in \cite{Heinze2016,Supanitsky2016,Berezinsky2016}.

	\paragraph{Astrophysical neutrinos ---} 
	We tested astrophysical neutrino models for the UHECR sources.
	One of the advantages of studying astrophysical neutrino models is that not only proton-dominant, but also mixed- or heavy-composition UHECR 
	models can be tested with IceCube.
	The results of the model tests are listed in Table\;\ref{table:CL2}, and the limits are shown in the lower panel of Fig.\;\ref{fig:MD}. 
	
	The AGN models relate the neutrino emission rates in each source with the observed photon fluxes using phenomenological parameters, 
	such as the baryon loading factor $\xi_{cr}$\;\cite{murase14} and the neutrino-to-$\gamma$-ray intensity ratio $Y_{\nu\gamma}$\;\cite{padovani15}. 
	As the neutrino flux scales linearly with these parameters, the limits can be interpreted as constraints on the parameters, as listed in Table\;\ref{table:CL2}.
	The observed UHECR generation rate around $10^{10}$ GeV	($\sim10^{44}$\;erg\;Mpc$^{-3}$\;yr$^{-1}$) requires the loading factor $\xi_{cr}$ to be around 3 and 100 
	for UHECR spectral indices $s=2.0$ and $2.3$, respectively \cite{murase14}. The current constraints on $\xi_{cr}$ are comparable or slightly below these required values. This
	indicates that AGN inner jets are less likely to be a major source of the UHECRs, regardless of the observed UHECR compositions.
	A consistent but weaker limit on these models is also obtained from an analysis searching for the neutrino signal excess in the direction of blazer populations \cite{Glusenkamp:2015jca}.
	Rapidly spinning pulsars may also be capable of accelerating nuclei to $10^{11}$\;GeV \cite{ke14}. They are also disfavored as UHECR sources if they have cosmological evolution stronger than SFR.
	As shown in Fig.\;\ref{fig:MD}, provided a flat neutrino spectrum in the UHECR source is assumed, astrophysical neutrino spectra are generally 
	predicted to be described by a hard power law\;\cite{muraseAGN}.
	These spectra continue up to a cutoff energy determined by the maximal acceleration energy of the source.
	Figure\;\ref{fig:contours} provides a generic constraint on these astrophysical fluxes as an exclusion region in the parameter space for $E^{-2}$ power-law neutrino flux normalization $\phi_0$ and spectral cutoff energy $E^{cut}_{\nu}$. 
	It indicates that $E^2\phi_0 \geq 6 \times 10^{-9}$ GeV cm$^{-2}$ s$^{-1}$ sr$^{-1}$ is disfavored for neutrino fluxes extending 
	above $10^{9}$\;GeV, such as the UHECR source models.  
	\paragraph{Differential limit ---} 
	A quasi-differential 90\%-CL limit is presented in Fig.~\ref{fig:differential} using the LLR method, considering the two observed events. 
	Each point on the solid line is the result of an independent hypothesis test for a decade-wide $E^{-1}$ power-law flux as a signal model, representing a 90\%-CL upper limit.
	The median null observation limit (sensitivity) is also presented.
	The limit for an $E^{-2}$ flux ($E_{\nu}^2\phi < 9.2\times10^{-9}$ GeV/cm$^2$\;s\;sr) 
	in the central 90\% energy region between $1.0\times10^6$ and $4.0\times10^9$ GeV is shown for reference.
	\paragraph{Summary ---}
	Analysis of IceCube data results in the largest exposure to date in search for the neutrino flux above $10^7$\;GeV up to $3 \times 10^{10}$\;GeV. 
	The non-observation of neutrino events with deposited energy larger than a few PeV in seven years of IceCube data places a serious constraint on cosmogenic and astrophysical neutrino models.
	The restrictions on the cosmological evolution of 
	UHECR sources and the model-dependent constraints on the source classes reported herein are the strongest constraints on the origin of 
	the highest-energy cosmic rays above the ankle achieved via neutrino astronomy.
	The detection of cosmogenic neutrinos from sources with weak or no evolution, and of heavy-composition UHECRs requires a larger scale detector. Cost-effective radio Askaryan neutrino detectors, such as ARA~\cite{ara16} or ARIANNA~\cite{arianna15}, therefore would be an important future option.  
	\begin{table}[bt]
		\begin{center}
			\begin{tabular}{lcrc}
				\hline
				\hline
				$\nu$ Model & Event rate & p-value & MRF\\
				 & per livetime & &\\
				\hline
				Kotera {\it et al.}~\cite{kotera2010} &    && \\
				SFR & $3.6^{+0.5}_{-0.8}$ & $22.3^{+10.8}_{-3.9}$\% & 1.44\\ 
				\hline
				Kotera {\it et al.}~\cite{kotera2010} &    && \\
				FRII  & $14.7^{+2.2}_{-2.7}$ &  $<$0.1\% &0.33\\ 
				\hline
				Aloisio {\it et al.}~\cite{SimProp15}&&&\\
				SFR &  $4.8^{+0.7}_{-0.9}$ &  $7.8^{+6.8}_{-1.8}$\% & 1.09\\
				\hline
				Aloisio {\it et al.}~\cite{SimProp15}&&&\\
				FRII  &  $24.7^{+3.6}_{-4.6}$ & $<$0.1\% & 0.20\\ 
				\hline
				Yoshida {\it et al.}~\cite{yoshida93} &    & & \\
				$m = 4.0,z_{max}=4.0$&  $7.0^{+1.0}_{-1.0}$ & $0.1^{+0.4}_{-0.1}$\%  &0.37\\
				\hline
				Ahlers {\it et al.}~\cite{ahlers2010} &    && \\
				best fit, 1 EeV&  $2.8^{+0.4}_{-0.4}$  & $9.5^{+6.5}_{-1.6}$\% &1.17\\
				\hline
				Ahlers {\it et al.}~\cite{ahlers2010} &    && \\
				best fit, 3 EeV & $4.4^{+0.6}_{-0.7}$  & $2.2^{+1.3}_{-0.9}$\% &0.66\\
				\hline
				Ahlers {\it et al.}~\cite{ahlers2010} &    && \\
				best fit, 10 EeV & $5.3^{+0.8}_{-0.8}$  & $0.7^{+1.6}_{-0.2}$\% &0.48\\				
				\hline
				\hline 
				\hline
			\end{tabular}
			\caption{Cosmogenic neutrino model tests: Expected number of events in 2426 days of effective livetime,
				p-values from model hypothesis test, and 90\%-CL model-dependent limits in terms of the model rejection factor (MRF)~\cite{MRF}, 
				defined as the ratio between the flux upper limit and the predicted flux.}
			\label{table:CL}
		\end{center}
	\end{table}
	\begin{table}[bt]
		\begin{center}
			\begin{tabular}{lcrc}
				\hline
				\hline
				$\nu$ Model & Event rate & p-value & MRF\\
				& per livetime & &\\
				\hline
				Murase~{\it et al.}~\cite{murase14} &    & & \\
				$s=2.3$, $\xi_{CR}$=100& $7.4^{+1.1}_{-1.8}$  &  $2.2^{+9.9}_{-1.4}$\% & 0.96 ($\xi_{CR}\leq$96)\\
				\hline
				Murase~{\it et al.}~\cite{murase14} &    & & \\
				$s=2.0$, $\xi_{CR}$=3&  $4.5^{+0.7}_{-0.9}$ &  $19.9^{+20.2}_{-9.2}$\% & 1.66 ($\xi_{CR}\leq$5.0)\\
				\hline
				Fang~{\it et al.}~\cite{ke14} &    && \\
				SFR&  $5.5^{+0.8}_{-1.1}$  & $7.8^{+14.4}_{-3.7}$\% &1.34\\
				\hline
				Fang~{\it et al.}~\cite{ke14} &    && \\
				uniform&  $1.2^{+0.2}_{-0.2}$  & $54.8^{+1.7}_{-2.7}$\% & 5.66\\
				\hline
				Padovani~{\it et al.}~\cite{padovani15} &    && \\
				$Y_{\nu\gamma}=0.8$&  $37.8^{+5.6}_{-8.3}$  & $<$0.1\% & 0.19 ($Y_{\nu\gamma}\leq$0.15)\\
				\hline
				\hline
			\end{tabular}
			\caption{Astrophysical neutrino model tests: Same as Table \ref{table:CL}. The flux normalization scales linearly for AGN models with the assumed 
				baryonic loading factor $\xi_{CR}$ for 	Murase~FSRQ (broad-line region)~\cite{murase14} or neutrino-to-$\gamma$ ratio $Y_{\nu \gamma}$ for Padovani BL\;Lac~\cite{padovani15} models. 
				A power-law proton UHECR
				spectrum with index $s$ is assumed in the FSRQ model.
				The corresponding 
				parameters for these models to explain the measured IceCube neutrino flux in TeV-PeV range\;\cite{HESEPRL} are excluded by more than 99.9\% CL.}
			\label{table:CL2}
		\end{center}
	\end{table}
	\begin{acknowledgments}
		
		We acknowledge the support from the following agencies:
		U.S. National Science Foundation-Office of Polar Programs,
		U.S. National Science Foundation-Physics Division,
		University of Wisconsin Alumni Research Foundation,
		the Grid Laboratory Of Wisconsin (GLOW) grid infrastructure at the University of Wisconsin - Madison, the Open Science Grid (OSG) grid infrastructure;
		U.S. Department of Energy, and National Energy Research Scientific Computing Center,
		the Louisiana Optical Network Initiative (LONI) grid computing resources;
		Natural Sciences and Engineering Research Council of Canada,
		WestGrid and Compute/Calcul Canada;
		Swedish Research Council,
		Swedish Polar Research Secretariat,
		Swedish National Infrastructure for Computing (SNIC),
		and Knut and Alice Wallenberg Foundation, Sweden;
		German Ministry for Education and Research (BMBF),
		Deutsche Forschungsgemeinschaft (DFG),
		Helmholtz Alliance for Astroparticle Physics (HAP),
		Research Department of Plasmas with Complex Interactions (Bochum), Germany;
		Fund for Scientific Research (FNRS-FWO),
		FWO Odysseus programme,
		Flanders Institute to encourage scientific and technological research in industry (IWT),
		Belgian Federal Science Policy Office (Belspo);
		University of Oxford, United Kingdom;
		Marsden Fund, New Zealand;
		Australian Research Council;
		Japan Society for Promotion of Science (JSPS);
		the Swiss National Science Foundation (SNSF), Switzerland;
		National Research Foundation of Korea (NRF);
		Villum Fonden, Danish National Research Foundation (DNRF), Denmark
		
	\end{acknowledgments}
	
		\clearpage
				\ifx \standalonesupplemental\undefined
		\setcounter{page}{1}
		\setcounter{figure}{0}
		\setcounter{table}{0}
		\fi
		
				\ifx \standaloneerratum\undefined
		\setcounter{page}{1}
		\setcounter{figure}{0}
		\setcounter{table}{0}
		\fi
		
		\newcolumntype{L}[1]{>{\arraybackslash}p{#1}}
		\newcolumntype{C}[1]{>{\centering\arraybackslash}p{#1}}
		\newcolumntype{R}[1]{>{\hfill\arraybackslash}p{#1}}
		
		\renewcommand{\thepage}{Erratum -- E\arabic{page}}
		\renewcommand{\figurename}{Erratum FIG.}
		\renewcommand{\tablename}{Erratum TABLE}

\section{Erratum: Constraints on ultra-high-energy cosmic ray sources from a search for neutrinos above 10 PeV with IceCube}
\vspace{5mm}

In a recent letter \cite{icecube2016}, we stated that two neutrino-induced events were detected.
The observed events were, because of their estimated energies, interpreted as background in the original analysis searching for neutrinos above 10 PeV.
One of two events was 
described as a particle shower with a deposited energy of $(7.7\pm2.0) \times 
10^5$ GeV.
Later investigation revealed that this event was a detector artifact caused by a spurious 
flash from the in-ice calibration laser during the warm-up period before a calibration run. We have updated the current analysis excluding all the runs overlapping with the laser warm-up period.
The total live time difference with the update is less than 0.5\%. 
The other neutrino-induced event, an upward going track with a 
deposited energy of $(2.6\pm0.3) \times 10^6$ GeV, is unaffected. 
A further search identified no other high-energy neutrino candidates affected by the calibration laser.


The atmospheric background-only hypothesis of detecting the 
one surviving event is rejected at $2.2\sigma$. The observed event is compatible with a generic 
astrophysical $E^{-2}$ power-law flux with a p-value of 86.4\% and the hypothesis that this event is of cosmogenic origin is rejected with a p-value of 2.2\%. 
%
The corresponding evaluation of representative models is given in Table \ref{table:CL} and \ref{table:CL2} as well as the model-dependent limits presented in Fig.\;\ref{fig:MD}. 
The quasi-differential limit and a model-dependent upper limit on an unbroken $E^{-2}$ power-law flux shown in Figure 4 of the original letter become stronger. An updated version of this plot can be found in Fig. \ref{fig:differential}.

An updated exclusion contour from a generic scanning of the parameter space for the source evolution function, $\varPsi_s(z)\propto (1+z)^m$, up to the maximum source extension 
in redshift $z\leq z_{\rm max}$, is shown in the upper panel of Fig.\;\ref{fig:contours}. The lower panel of Fig.\;\ref{fig:contours} provides a generic constraint on these astrophysical fluxes as an exclusion region in the parameter space of $E^{-2}$ power-law neutrino flux normalization $\phi_0$ and spectral cutoff energy $E^{cut}_{\nu}$. 
\begin{table}[b]
	\begin{center}
		\begin{tabular}{lcrc}
			\hline
			\hline
			$\nu$ Model & Event rate & p-value & MRF\\
			& per livetime & &\\
			\hline
			Kotera {\it et al.} &    && \\
			SFR & $3.6^{+0.5}_{-0.8}$ & $6.0^{+2.9}_{-1.0}$\% & 1.04\\ 
			\hline
			Kotera {\it et al.} &    && \\
			FRII  & $14.7^{+2.2}_{-2.7}$ &  $<$0.1\% &0.23\\ 
			\hline
			Aloisio {\it et al.}&&&\\
			SFR &  $4.8^{+0.7}_{-0.9}$ &  $3.2^{+2.8}_{-0.7}$\% & 0.80\\
			\hline
			Aloisio {\it et al.}&&&\\
			FRII  &  $24.7^{+3.6}_{-4.6}$ & $<$0.1\% & 0.15\\ 
			\hline
			Yoshida {\it et al.}&    & & \\
			$m = 4.0,z_{max}=4.0$&  $7.0^{+1.0}_{-1.0}$ & $0.1^{+0.4}_{-0.1}$\%  &0.43\\
			\hline
			Ahlers {\it et al.}&    && \\
			best fit, 1 EeV&  $2.8^{+0.4}_{-0.4}$  & $13.4^{+9.2}_{-2.2}$\% &1.33\\
			\hline
			Ahlers {\it et al.}&    && \\
			best fit, 3 EeV & $4.4^{+0.6}_{-0.7}$  & $3.2^{+1.8}_{-1.4}$\% &0.76\\
			\hline
			Ahlers {\it et al.}&    && \\
			best fit, 10 EeV & $5.3^{+0.8}_{-0.8}$  & $1.1^{+2.5}_{-0.3}$\% &0.63\\				
			\hline
			\hline 
			\hline
		\end{tabular}
		\caption{Cosmogenic neutrino model tests: Expected number of events in the effective livetime,
			p-values from model hypothesis test, and 90\%-CL model-dependent limits in terms of the model rejection factor (MRF). See the caption of the original letter for full citations.}
		\label{table:CL}
	\end{center}
\end{table}
\begin{table}
	\begin{center}
		\begin{tabular}{lcrc}
			\hline
			\hline
			$\nu$ Model & Event rate & p-value & MRF\\
			& per livetime & &\\
			\hline
			Murase~{\it et al.}&    & & \\
			$s=2.3$, $\xi_{CR}$=100& $7.4^{+1.1}_{-1.8}$  &  $0.3^{+1.3}_{-0.2}$\% & 0.62 ($\xi_{CR}\leq$62)\\
			\hline
			Murase~{\it et al.}&    & & \\
			$s=2.0$, $\xi_{CR}$=3&  $4.5^{+0.7}_{-0.9}$ &  $4.8^{+4.9}_{-2.2}$\% & 1.32 ($\xi_{CR}\leq$4.0)\\
			\hline
			Fang~{\it et al.}&    && \\
			SFR&  $5.5^{+0.8}_{-1.1}$  & $1.6^{+3.0}_{-0.8}$\% &0.88\\
			\hline
			Fang~{\it et al.}&    && \\
			uniform&  $1.2^{+0.2}_{-0.2}$  & $78.2^{+2.4}_{-3.9}$\% & 4.0\\
			\hline
			Padovani~{\it et al.}&    && \\
			$Y_{\nu\gamma}=0.8$&  $37.8^{+5.6}_{-8.3}$  & $<$0.1\% & 0.12 ($Y_{\nu\gamma}\leq$0.13)\\
			\hline
			\hline
		\end{tabular}
		\caption{Astrophysical neutrino model tests: Same as Table \ref{table:CL}. See the caption of the original letter for full citations.}
		\label{table:CL2}
	\end{center}
\end{table}
\begin{figure}
	\begin{center}
		\includegraphics[width=3.4in]{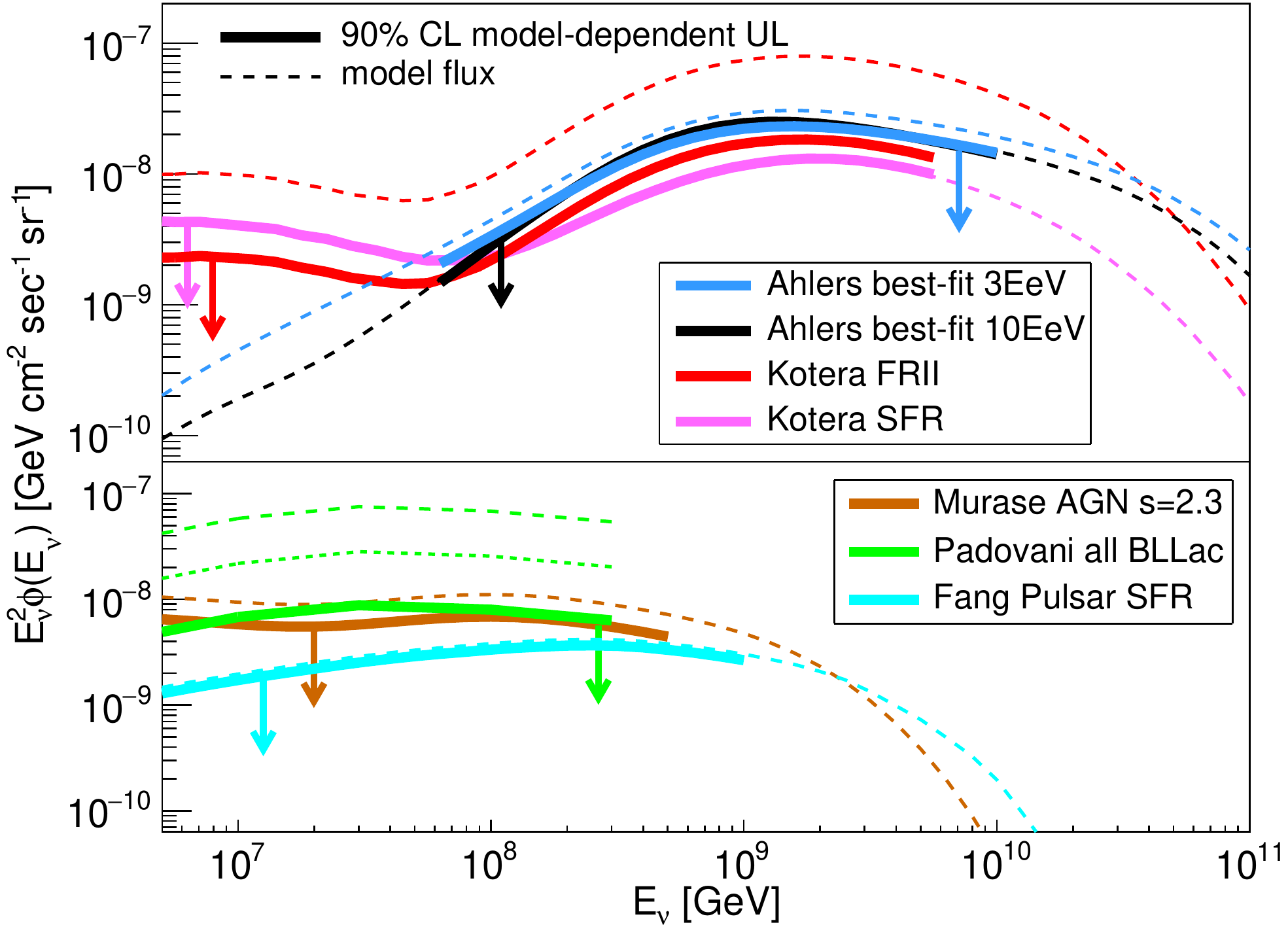}
	\end{center}
	\caption{
		Model-dependent 90\% confidence-level limits (solid lines) for cosmogenic and astrophysical neutrino predictions.
		The range of limits indicates the central 90\% energy region. See the caption of the original letter for full details.
	}
	\label{fig:MD}
\end{figure}
\begin{figure}
	\begin{center} 				
		\includegraphics[width=3.2in]{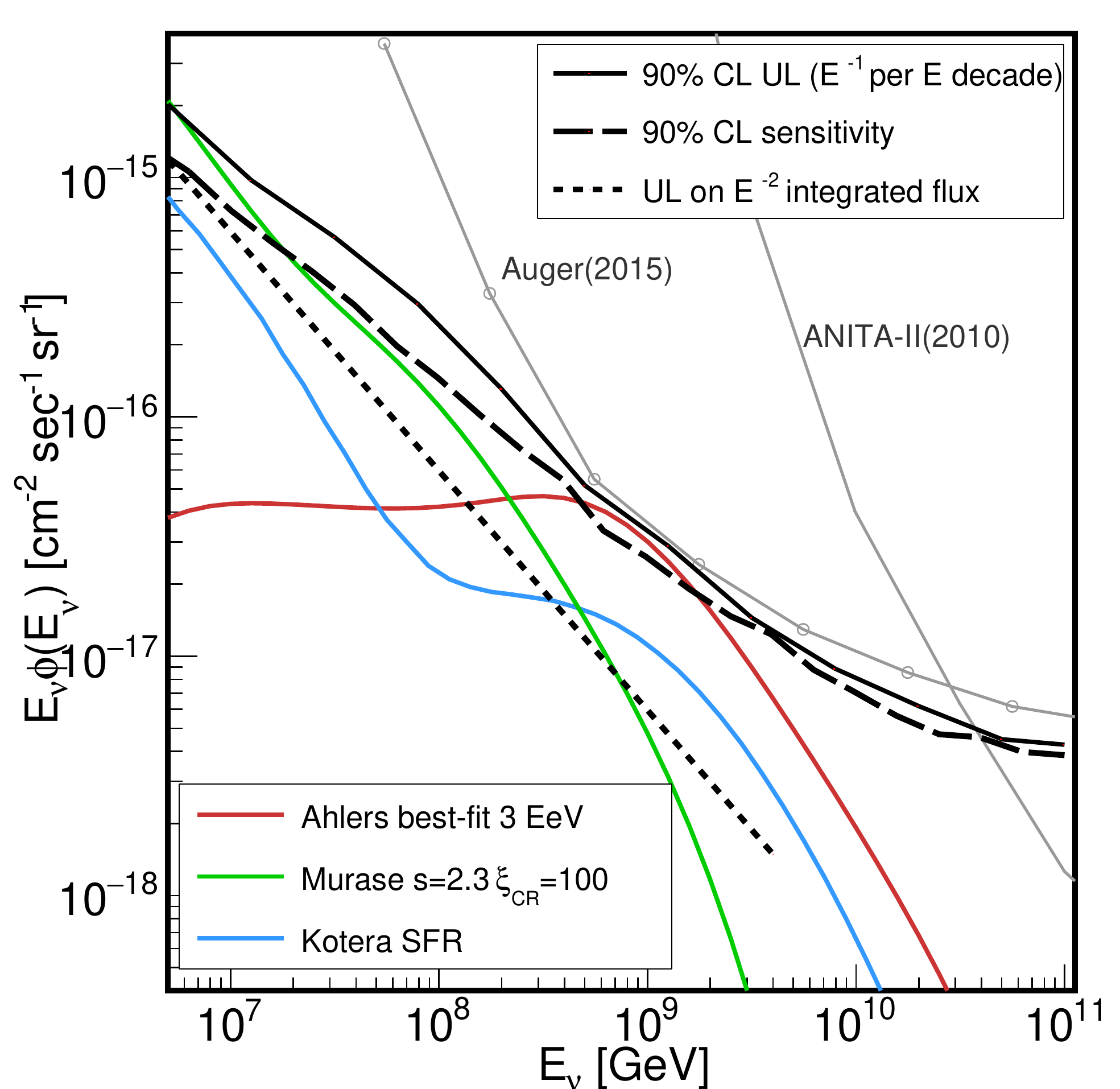}
	\end{center}
	\caption{All-flavor-sum neutrino flux quasi-differential 90\%-CL upper limit on one energy decade $E^{-1}$ flux windows (solid line). 
		A model-dependent upper limit on an unbroken $E^{-2}$ power-law flux from the current analysis 
		($E_{\nu}^2\phi < 5.9\times10^{-9}$ GeV/cm$^2$\;s\;sr) is also shown (dotted line). See the caption of the original letter for full details.
	}
	\label{fig:differential}
\end{figure}	

\begin{figure}
	\begin{center}
		\includegraphics[height=1.5in]{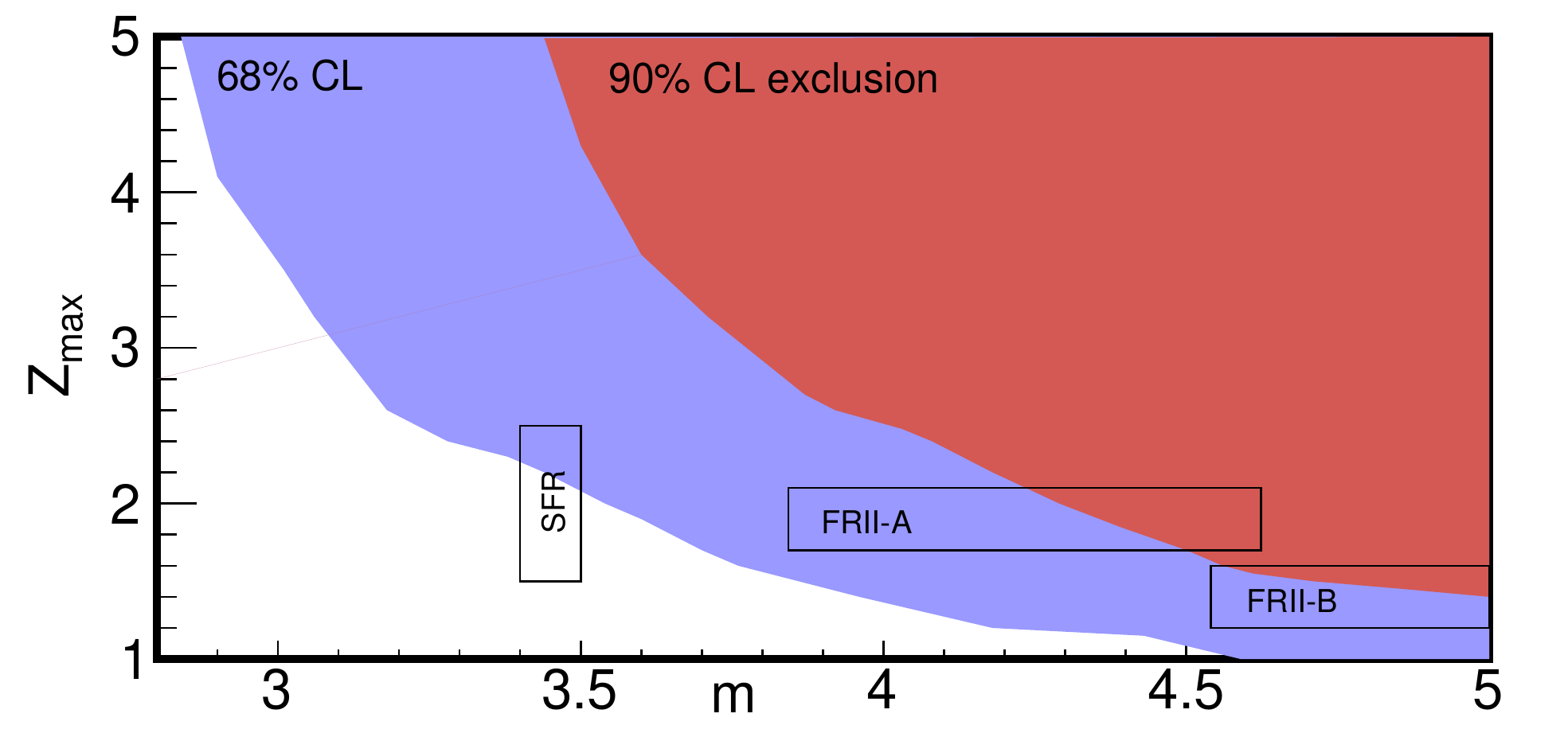}\\
		\includegraphics[height=1.5in]{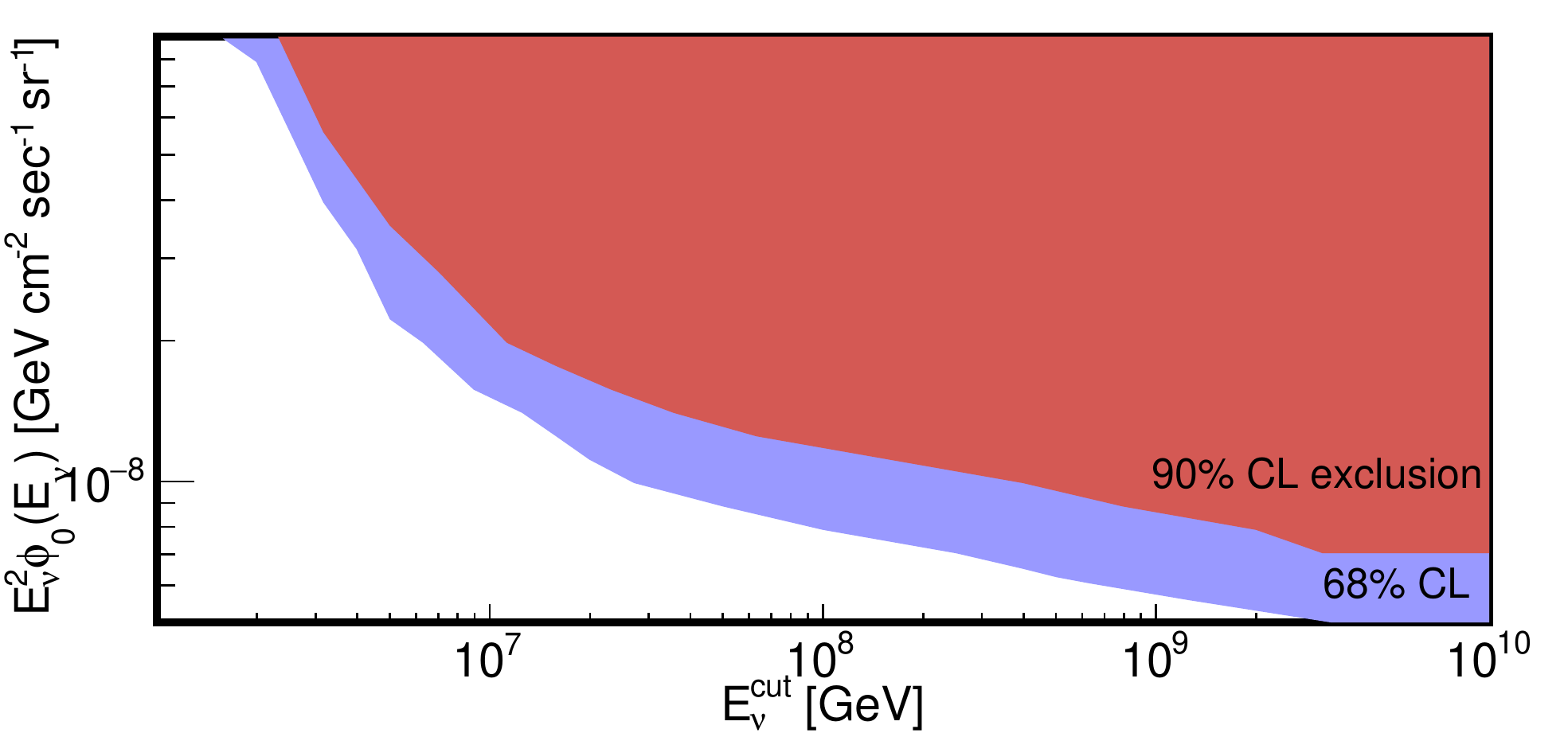}
	\end{center}	
	\caption{Constraints on UHECR source evolution model and all flavor $E^{-2}$ power-law 
		flux model parameters. The colored areas represent parameter space excluded by the current analysis.
		(Top) Cosmogenic flux parameters $m$ and $z_{max}$ of UHECR-source cosmological evolution 
		function of the form $\psi_s(z) \propto (1+z)^m$.
		(Bottom) 	
		Upper limits on $E^{-2}$ power-law neutrino flux normalization $\phi_0$ and spectral cutoff energy $E^{cut}_{\nu}$. 
		See the caption of the original letter for full details.
	}
	\label{fig:contours}
\end{figure}


		\clearpage
		
		\ifx \standalonesupplemental\undefined
		\setcounter{page}{1}
		\setcounter{figure}{0}
		\setcounter{table}{0}
		\fi
		
		\newcolumntype{L}[1]{>{\arraybackslash}p{#1}}
		\newcolumntype{C}[1]{>{\centering\arraybackslash}p{#1}}
		\newcolumntype{R}[1]{>{\hfill\arraybackslash}p{#1}}
		
		\renewcommand{\thepage}{Supplementary Methods and Tables -- S\arabic{page}}
		\renewcommand{\figurename}{SUPPL. FIG.}
		\renewcommand{\tablename}{SUPPL. TABLE}
		
		\vspace{50mm}
		
		\section{Supplementary Methods and Tables}
			\vspace{5mm}

		\section*{Data Selection}
		In this section, the event selection used in this study is described.
		%
		The event selection is designed to retain a large part of the simulated cosmogenic neutrino sample from \cite{ahlers2010} while eliminating the simulated background.
		The background events in this analysis are induced by downward-going muon bundles and atmospheric neutrinos from any direction produced in air showers.
		These background events are expected to have softer spectra compared with those of the signal cosmogenic and astrophysical neutrino events.

		\paragraph{Online EHE Filter:}
		As an input of the analysis, the sample selected by an extremely high energy (EHE) filter is used.
		The EHE filter eliminates a large amount of atmospheric-muon-induced events with low energy, typically less than a few TeV, by requiring the total number of photo-electrons (NPE) to be greater than 1,000 photo-electrons (p.e.).
		This filtering process is performed at the South Pole, and the resulting samples are sent to the data warehouse in the northern hemisphere via satellite.
		\paragraph{Offline EHE Cut:}
		Additional hit cleaning is applied offline to the EHE filter sample, in order to remove coincident atmospheric muons and PMT noise, for recalculating NPE.
		More than 25,000 p.e. in the offline NPE and more than 100 optical sensors with non-zero photo-electrons are required to further reduce the background-induced events.
		The zenith angle of these events is reconstructed using a chi-squared fit, assuming a simple track hypothesis.
		
\begin{figure}
			\centering
			\subfloat[0.28\textwidth][Rate of simulated cosmogenic neutrino events in 2426 days.]{
				\includegraphics[width=7.5cm]{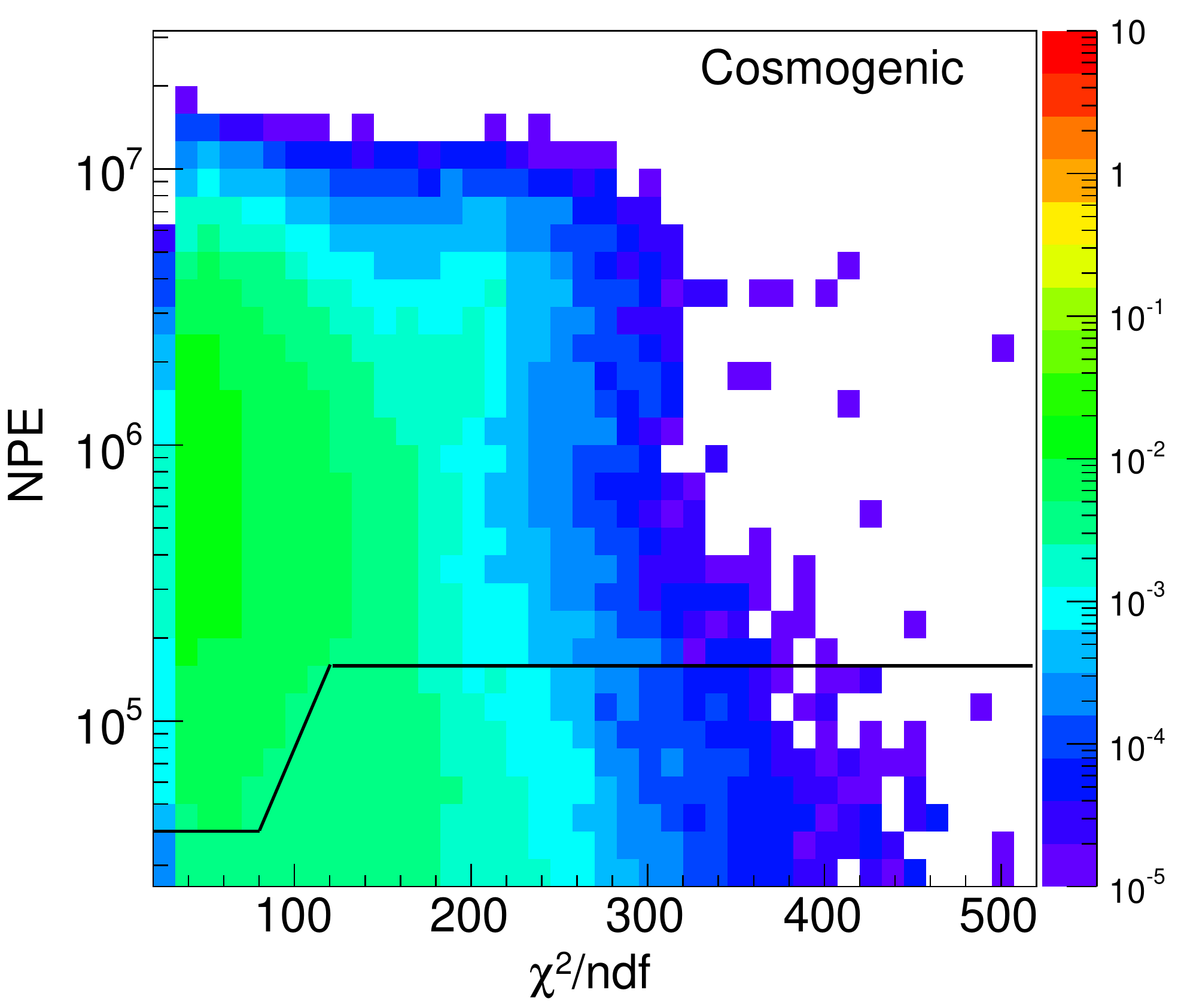}
			}
			
			\subfloat[0.28\textwidth][Rate of simulated air shower events in 2426 days.]{
				\includegraphics[width=7.5cm]{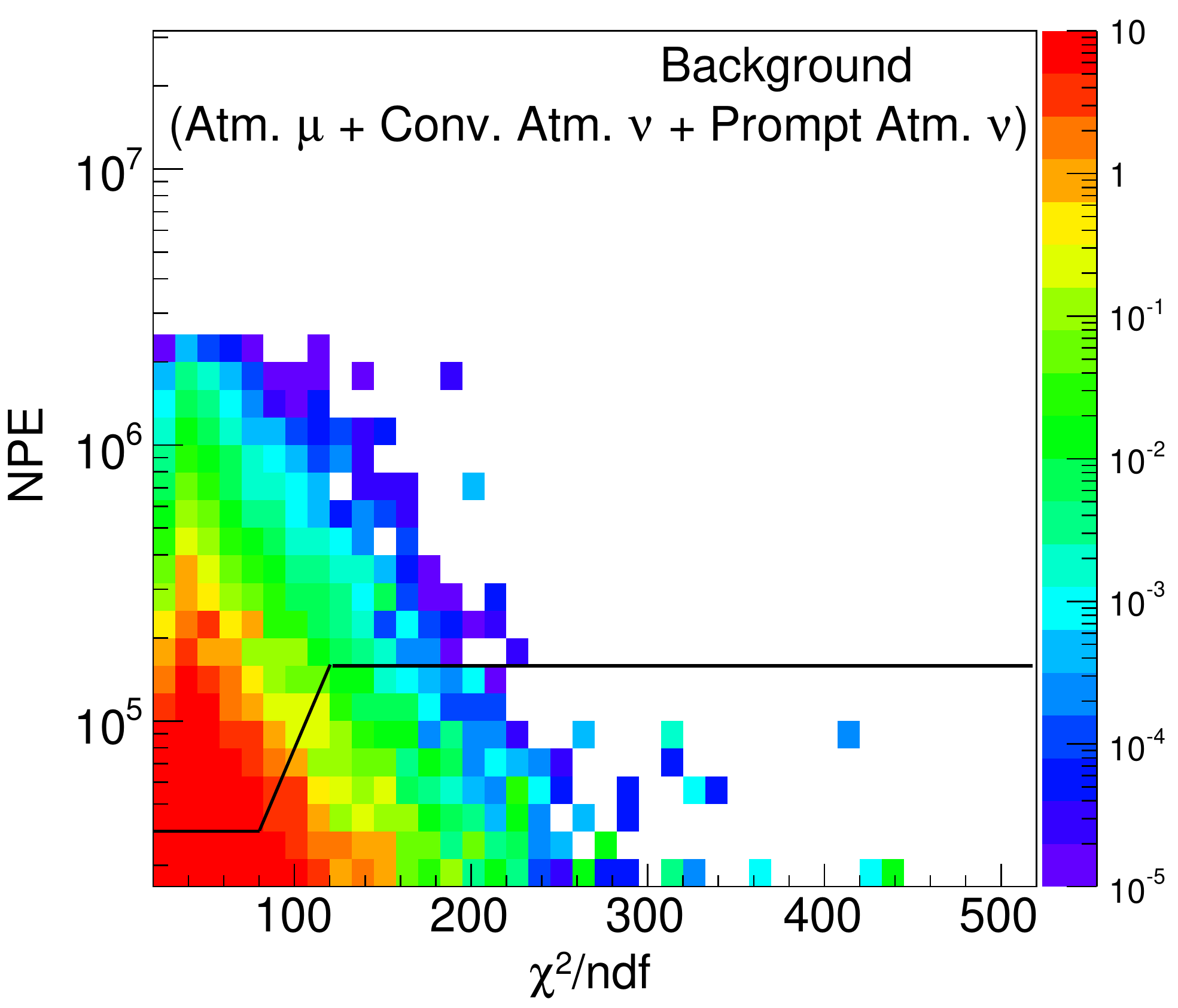}
			}
			\caption{Event number distributions before the track quality cut of the sample including all three flavors of neutrinos as a function of NPE and $\chi^2_{\rm track}/ndf$. The solid line in each panel indicate the track quality selection criteria for which only the events above the lines are retained. Event distributions from different detector configurations are summed. The selection criteria are constant for the samples taken in different data-recording periods.}
			\label{fig:L3_cut}
		\end{figure}

		\paragraph{Track Quality Cut:}
		These data are subject to a ``track quality'' cut, which aims to remove atmospheric neutrinos, particularly the prompt neutrinos with a large theoretical uncertainty, and atmospheric muons whose directions are not reliably reconstructed.
		This was achieved by setting tighter criteria for the events with a larger reduced chi-squared ($\chi^2_{\rm track}/ndf$) associated with the zenith angle reconstruction.
		Suppl. Fig. \ref{fig:L3_cut} shows the event distributions as a function of the $\chi^2_{\rm track}/ndf$ value and NPE.
		The track quality selection of the NPE threshold as a function of reduced $\chi^2$ is shown by the solid line in Suppl. Fig.~\ref{fig:L3_cut}.
		
\begin{figure}
			\centering
			\subfloat[0.28\textwidth][Rate of simulated cosmogenic neutrino events in 2426 days.]{
				\includegraphics[width=7.5cm]{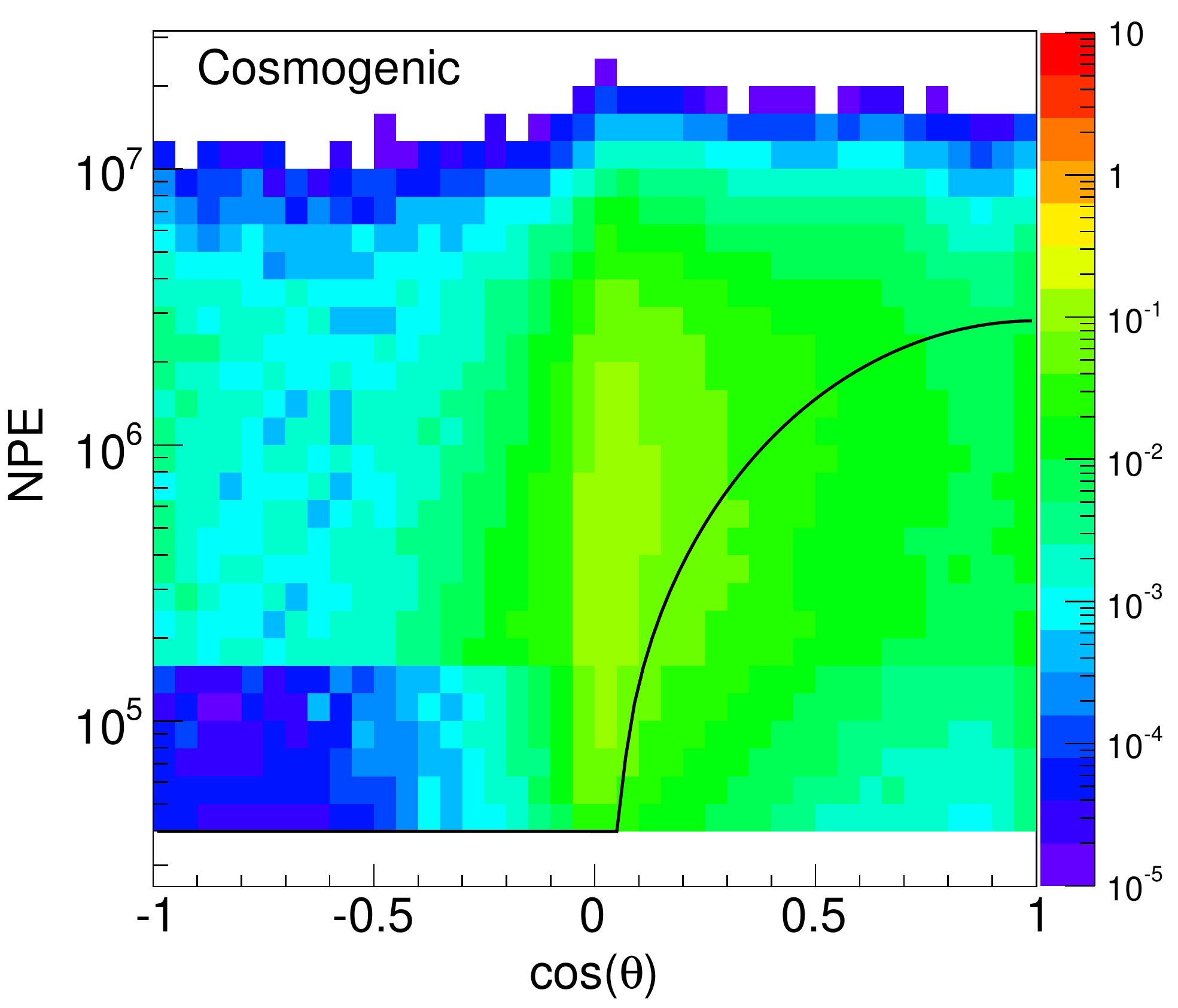}
			}
			
			\subfloat[0.28\textwidth][Rate of simulated air shower events in 2426 days.]{
				\includegraphics[width=7.5cm]{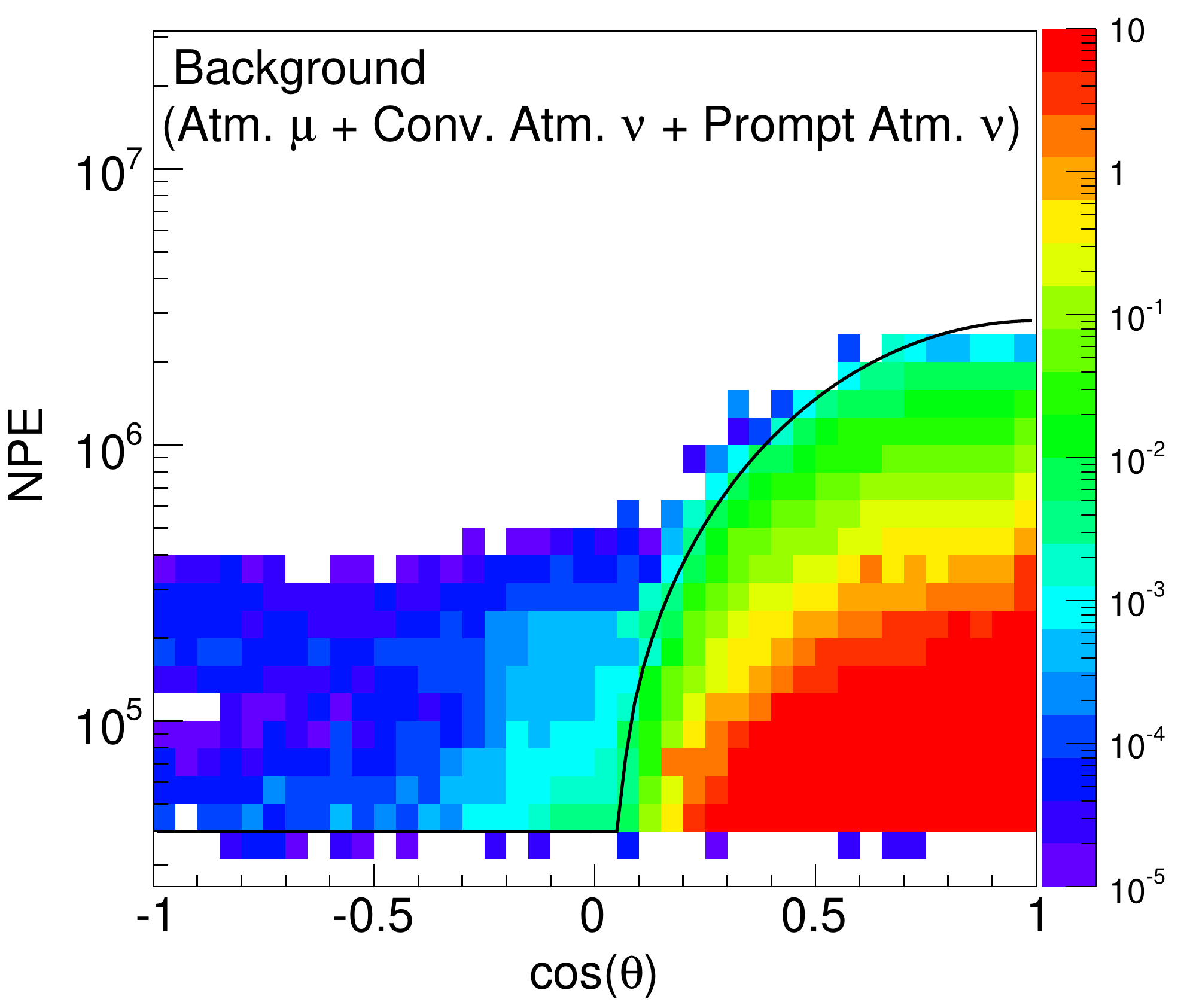}
			}
			\caption{Event number distributions before the muon bundle cut, including all three flavors of neutrinos as a function of NPE and $cos(\theta)$. The solid line in each panel indicate the muon bundle selection criteria for which only the events above the lines are retained. Event distributions from different detector configurations are added. The selection criteria are constant for the samples taken in different data-recording periods.}
			\label{fig:L4_cut}
		\end{figure}

		\paragraph{Muon Bundle Cut:}
		The ``muon bundle'' cut is developed to remove the remaining well-reconstructed atmospheric muons in the downward-going directions.
		It also ensures that there are no IceTop hits from cosmic-ray air showers associated with events near the final level.
		Suppl. Fig. \ref{fig:L4_cut} shows the event distributions as a function of the cosine of the reconstructed zenith angle ($cos(\theta)$) and NPE.
		The muon bundle cut criteria of the NPE threshold as a function of $cos(\theta)$ are shown by the solid line in Suppl. Fig.~\ref{fig:L4_cut}.

		Suppl. Tab.~\ref{tab:cut_progression} shows results of the data selection process. The background from cosmic-ray air showers is reduced by a factor of approximately $1.2 \times 10^{10}$, while 42\% of the neutrinos from a hypothetical cosmological neutrino flux with respect to the EHE online filter are expected to be retained. 
		
		\begin{table*}[b]
			\centering
			\begin{tabular}{|c|c|c|c|}
				\hline
				Cut level & atmos. muons & atmos. neutrinos & signal cosmogenic neutrinos \cite{ahlers2010}\\
				& number in 2426-d & number in 2426-d & fraction surviving (\%) \\
				\hline
				{ Online EHE Filter}
				& $1.7 \times 10^{8}$ & $5.0 \times 10^2$& 100 \\
				\hline
				{ Offline EHE Cut} 
				& $3.2 \times 10^{5}$& 1.2&  74\\
				\hline
				{ Track Quality Cut}
				& $8.0 \times 10^{3}$& $8.3 \times 10^{-2}$&  61\\
				\hline
				{ Muon bundle Cut}
				&$2.1 \times 10^{-2}$ &$4.3 \times 10^{-2}$ & 42 \\
				\hline
			\end{tabular}
			\caption{Rates and fractions of simulated data surviving by type as a function of the level of selection applied. Efficiencies are with respect to the online EHE filter.}	\label{tab:cut_progression}
		\end{table*}
			
	\end{document}